# Chaotic Analog-to-Information Conversion: Principle and Reconstructability with Parameter Identifiability

Feng Xi, Sheng Yao Chen and Zhong Liu
Department of Electronic Engineering
Nanjing University of Science & Technology
Nanjing, Jiangsu 210094
The People's Republic of China
E-mails: xf_njust@yahoo.com.cn, chen_shengyao@163.com, eezliu@mail.njust.edu.cn

**Abstract**：This paper proposes a chaos-based analog-to-information conversion system for the acquisition and reconstruction of sparse analog signals. The sparse signal acts as an excitation term of a continuous-time chaotic system and the compressive measurements are performed by sampling chaotic system outputs. The reconstruction is realized through the estimation of the sparse coefficients with principle of chaotic parameter estimation. With the deterministic formulation, the analysis on the reconstructability is conducted via the sensitivity matrix from the parameter identifiability of chaotic systems. For the sparsity-regularized nonlinear least squares estimation, it is shown that the sparse signal is locally reconstructable if the columns of the sparsity-regularized sensitivity matrix are linearly independent. A Lorenz system excited by the sparse multitone signal is taken as an example to illustrate the principle and the performance.

# I. Introduction

With modern signal processing firmly rooted in digital computation, efficient conversion from analog signal to discrete one is of fundamental importance. In recent years, the demand to acquire data at ever increasing bandwidths has imposed a burden on traditional analog-to-digital converters that rely on the Shannon-Nyquist sampling theorem.

The desire to circumvent the Shannon-Nyquist limitation has prompted a new signal acquisition framework, compressive sampling (CS) or compressive sensing [Donoho, 2006; Candès *et al*., 2006; Candès & Tao, 2006]. With linear random projection, CS admits sparse signals to be represented by a lower rate signal. This theory has spawned a number of new sub-Nyquist sampling structures (also called analog-to-information conversion, A2I) including the random sampling [Laska *et al*., 2006], random demodulator [Laska *et al*., 2007; Tropp *et al*., 2010], random filter [Tropp *et al*., 2006], modulated wideband converter [Mishali & Eldar, 2010], and others. These structures try to implement the linear random measurements in CS theory and sample a wide swath of bandwidth at a rate significantly lower than twice its bandwidth. The reconstruction of the sparse signals is usually conducted through the convex optimization [Yin *et al*., 2008; Becker *et al*., 2011] or greedy iteration algorithms [Tropp & Gilbert, 2007; Needell & Tropp, 2008]. When the measurement satisfies the restricted isometry property (RIP) [Candès & Tao, 2005], the sparse signal can be exactly reconstructed from its sub-Nyquist samples.

Recently, we proposed a chaotic compressive sampling (ChaCS) in [Liu *et al*.,

2012], which performs low-rate sensing of discrete signals through chaotic projection. The compressive measurements are acquired by downsampling the outputs of the discrete-time chaotic system excited by the data to be compressed. Since ChaCS measurements are nonlinear, the compressive structure and the signal reconstruction are very different from the previous linear CS systems. ChaCS has several advantages over linear CS, including simple implementation structure, security of measurement data, reproductivity of the measurement system at a remote agent. However, the nonlinearity of the measurements increases the complexity of the reconstruction. What is important is that there are no explicit conditions, such as RIP in the linear CS, to guarantee the successful reconstruction of the sparse signal.

The purposes of this paper are twofold. The first one is to develop a sub-Nyquist sampling structure of the analog signals on the basis of ChaCS in [Liu *et al.*, 2012]. The second one is to derive sufficient reconstruction conditions of the sampled signals from the sub-Nyquist samples. The generalization of ChaCS to case of analog signals is shown in Fig.1 and is termed as chaotic analog-to-information (ChaA2I) conversion. As seen, the analog signal to be sampled is acted as an excitation term of a continuous-time chaotic system, and the sub-Nyquist samples are generated by sampling the output of the chaotic system. The continuous chaotic systems play a role similar to random spreading in A2I with random demodulation [Laska *et al.*, 2007]. Taking the excitation as a sparse signal, the reconstruction of the sparse coefficients can be implemented through the estimation of the excitation coefficients with principle of the parameter estimation. For sparse signals, a sparsity-regularized nonlinear least squares algorithm is proposed to reconstruct the sparse coefficients.

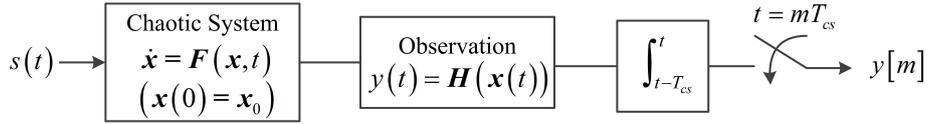

Fig.1 The structure of the ChaA2I converter.

An essential problem in the ChaCS or the proposed ChaA2I is whether the sub-Nyquist samples contain information enough to reconstruct the sparse signals. In linear CS theory, this is a well-developed area. If the measurement matrix satisfies the RIP condition, the sparse signal can be exactly reconstructed. The RIP condition will be guaranteed with high probability if the measurements are taken from some random distributions [Rudelson & Vershynin, 2008]. Different from the random measurement, the proposed scheme is deterministic! From a system point of view, the reconstruction is a problem to estimate the sparse coefficient parameters from sub-Nyquist samples. Then the reconstructability of the sparse signals is equivalent to the identifiability of the parameters from the sub-Nyquist samples. The topic of the parameter identifiability is well established in mathematical modeling within science and engineering [Jacquez & Greif, 1985; Zak et al., 2003; Rodriguez-Fernandez et al., 2006; Raue et al., 2009]. For our problem, the identifiability is studied by computing the correlation matrix of the sensitivity matrix [Zak et al., 2003] of the sparsity-regularized nonlinear least squares. It is found that the parameters are identifiable if the maximum absolute value of the off-diagonal elements of the correlation matrix is less than 1. With the analysis of the parameter identifiability, the appropriate sampling interval of the ChaA2I system can be determined, such that the sparse coefficient parameters are guaranteed to be identifiable.

The remainder of this paper is organized as follows. In Section II we introduce the signal model and problem. Section III describes the proposed ChaA2I and signal reconstruction. We conduct the analysis on the reconstructability in Section IV. Numerical results on the reconstructability and reconstruction performance are presented in Section V. Some conclusions are drawn in Section VI.

## II. Signal Model and Problem Statement

In the present work we are concerned with discrete acquisition of a real-valued, continuous-time signal $s(t)$ with observation interval $[0,1]$. It is assumed that $s(t)$ is represented in harmonical basis by

$$s(t) = \sum_{k=1}^{B} \alpha_k \psi_k(t) \tag{1}$$

where $\alpha_k$ ($k=1,2,\cdots,B$) are real-valued Fourier coefficients, $\psi_k(t) = \cos(2\pi k t)$ for $1 \leq k \leq B/2$ and $\psi_k(t) = \sin(2\pi(k - B/2)t)$ for $B/2+1 \leq k \leq B$ are real Fourier basis functions. With (1), it is implicitly assumed that the signal $s(t)$ is bandlimited with the bandwidth $B/2$. Let $\boldsymbol{\alpha} = [\alpha_1, \ldots, \alpha_B]^T$ and $\boldsymbol{\Psi}(t) = [\psi_1(t), \psi_2(t), \cdots, \psi_B(t)]$ be Fourier coefficient vector and Fourier basis vector, respectively. Then $s(t) = \boldsymbol{\Psi}(t)\boldsymbol{\alpha}$. For a sparse $s(t)$, the number $W$ of nonzero coefficients of $\boldsymbol{\alpha}$, $W = \|\boldsymbol{\alpha}\|_0$, is much smaller than $B$. The signal $s(t)$ is said to be $W$-sparse.

In the discrete-time acquisition of $s(t)$, the Shannon-Nyquist sampling theorem states that it is enough to represent $s(t)$ by $B$ samples of $s(t)$ per second with Nyquist sampling interval $T_{nq} = 1/B$. In contrast to the direct sampling, the recent CS theory indirectly samples the sparse $s(t)$ by $M$ indirect samples of $s(t)$ per second with $W < M < B$. The indirect samples are obtained by firstly randomizing the sparse signal $s(t)$ through a random operator $\varphi(\cdot): \mathbb{R} \to \mathbb{R}$ and then sampling the randomized output $\varphi(s(t))$ with the sampling interval $T_{cs} > T_{nq}$.

In this paper, we try to develop a random operator with chaotic system and design a chaotic analog-to-information conversion. In particular, the reconstructable condition of the sparse signal from its sub-Nyquist measurements is established.

## III. ChaA2I Converter

In this section, we introduce the ChaA2I converter proposed in this paper. Firstly, the structure of the ChaA2I converter is given. Then the reconstruction problem is discussed.

### A. Structure of ChaA2I Converter

The structure of the proposed ChaA2I converter is given in Fig.1, where the sparse signal $s(t)$ acts as the excitation to the chaotic system. With random-like behavior of chaotic system, the chaotic system generates the "randomized" output. Assume that the chaotic system is a $d$-dimensional continuous-time dynamic system

$$\dot{\mathbf{x}} = \mathbf{F}(\mathbf{x}, t) \quad (\mathbf{x}(0) = \mathbf{x}_0) \tag{2}$$

where $\mathbf{x} = [x_1, x_2, \ldots, x_d]^T \in \mathbb{R}^d$ and $\mathbf{F}(\mathbf{x}, t) = [f_1(\mathbf{x}, t), f_2(\mathbf{x}, t), \ldots, f_d(\mathbf{x}, t)]^T \in \mathbb{R}^d$ denote the state vector and the nonlinear field vector function, respectively. Let $\mathbf{x}_0$ be the initial state of the system. Without loss of generality, it is assumed that the excitation signal $s(t)$ is added to the state $x_1$. Then the autonomous chaotic system (2) becomes the following non-autonomous one

$$\dot{\mathbf{x}} = \begin{bmatrix} f_1(\mathbf{x},t) + \mu s(t) \\ f_2(\mathbf{x},t) \\ \vdots \\ f_d(\mathbf{x},t) \end{bmatrix} = \begin{bmatrix} f_1(\mathbf{x},t) + \mu \mathbf{\Psi}(t)\boldsymbol{\alpha} \\ f_2(\mathbf{x},t) \\ \vdots \\ f_d(\mathbf{x},t) \end{bmatrix} \quad (\mathbf{x}(0) = \mathbf{x}_0) \quad (3)$$

where $s(t)$ is represented as the basis representation form given in (1). The parameter $\mu > 0$ controls the coupling strength such that it does not destroy the chaotic behaviors of the system (3). Otherwise, (3) is acting as a linear or nonlinear filter and it is not enough to randomize the signal $s(t)$. For chaotic systems, the system output is random-like and is similar to that produced by random demodulation [Laska *et al.*, 2007].

With the non-autonomous chaotic system (3), we acquire the "randomized" output containing the information of the excitation signal $s(t)$. Then we can observe and sample the output of the chaotic system to get the sub-Nyquist measurements. Let $\mathbf{H}(\cdot): \mathbb{R}^d \to \mathbb{R}$ be an observation function which maps the $d$-dimensional state vector to a real-valued observation signal $y(t)$

$$y(t) = \mathbf{H}(\mathbf{x}(t;\mathbf{x}_0,\boldsymbol{\alpha})) \quad (4)$$

where $\mathbf{x}(t;\mathbf{x}_0,\boldsymbol{\alpha})$ denotes the state of system (3) at time $t$ with the initial state $\mathbf{x}_0$ and the excitation signal $s(t) = \mathbf{\Psi}(t)\boldsymbol{\alpha}$. Then $y(t)$ is lowpass filtered and sampled to output the discrete-time measurements. In Fig.1, an integrator is used to implement the lowpass filtering. Thus, we get a sequence of measurements $\mathbf{y} = [y_1, y_2, \ldots, y_M]^T$ with each measurement $y_m$ represented as

$$y_m = \int_{(m-1)T_{cs}}^{mT_{cs}} y(t)\,dt \qquad (m=1,\ldots,M) \tag{5}$$

where $T_{cs}$ denotes the sampling interval, and $M = \lfloor 1/T_{cs} \rfloor$. In practice, the lowpass filter and the sampler can be replaced by a standard A/D converter, which will simplify the implementation of the ChaA2I converter.

With the "randomization" provided by the chaotic system, the sparse signal $s(t)$ is mapped to an $M$-dimensional vector $\mathbf{y}$. If $T_{cs} > T_{nq}$, $M < B$. In this sense, the sparse signal $s(t)$ is compressed and $\mathbf{y}$ is the sub-Nyquist samples of $s(t)$. For an appropriate selection of $T_{cs}$, it is expected that the measurement $\mathbf{y}$ contains the information enough to reconstruct the sparse signal $s(t)$.

**B. Signal Reconstruction**

The reconstruction of the sparse signal $s(t)$ is equivalent to estimate the coefficient vector $\boldsymbol{\alpha}$. According to the CS theory, the sparse vector can be reconstructed by solving the $l_0$-norm minimization problem. For our case, we have the following $l_0$-norm minimization problem

$$\begin{cases} \min_{\bar{\boldsymbol{\alpha}}} \ \|\bar{\boldsymbol{\alpha}}\|_0 \\ \text{s.t.} \ \ y_m = \mathcal{H}_m(\mathbf{x}_0, \bar{\boldsymbol{\alpha}}) \quad m=1,2,\ldots,M \end{cases} \tag{6}$$

where

$$\mathcal{H}_m(\mathbf{x}_0, \bar{\boldsymbol{\alpha}}) = \int_{(m-1)T_{cs}}^{mT_{cs}} \mathbf{H}(\mathbf{x}(t;\mathbf{x}_0, \bar{\boldsymbol{\alpha}}))\,dt \tag{7}$$

is a sequence of measurements for the coefficient vector $\bar{\boldsymbol{\alpha}}$. With the constraint in (6), we can obtain an estimate of $\boldsymbol{\alpha}$ by solving (6). For the linear CS measurements, the constraint of (6) is linear on $\bar{\boldsymbol{\alpha}}$ and there are two classes of efficient algorithms to solve the $l_0$-norm minimization problem. One is the class of the greedy iteration algorithms [Tropp & Gilbert, 2007; Needell & Tropp, 2008] and another is the class of the $l_0$-norm relaxation algorithms [Yin *et al*., 2008; Becker *et al*., 2011] (see [Fornasier, 2010] for a review). Because of nonlinearity of chaotic systems, (6) is a non-convex nonlinear optimization problem. These algorithms are not directly feasible to (6).

We note that the Fourier representation vector $\boldsymbol{\alpha}$ is the excitation parameter of the non-autonomous chaotic system (3). Then, the sparse signal reconstruction or sparse vector estimation $\bar{\boldsymbol{\alpha}}$ of $\boldsymbol{\alpha}$ can be thought as the estimation of the model parameter of the non-autonomous system (3). Considering the sparsity of $\boldsymbol{\alpha}$, we define a sparsity-regularized cost function for the parameter estimation problem

$$\mathcal{L}_\lambda(\mathbf{x}_0, \bar{\boldsymbol{\alpha}}) = \|\mathbf{y} - \mathcal{H}(\mathbf{x}_0, \bar{\boldsymbol{\alpha}})\|_2^2 + \lambda \|\bar{\boldsymbol{\alpha}}\|_0 \tag{8}$$

where $\lambda$ is the regularization parameter and

$$\mathcal{H}(\mathbf{x}_0, \bar{\boldsymbol{\alpha}}) = [\mathcal{H}_1(\mathbf{x}_0, \bar{\boldsymbol{\alpha}}), \mathcal{H}_2(\mathbf{x}_0, \bar{\boldsymbol{\alpha}}), \ldots, \mathcal{H}_M(\mathbf{x}_0, \bar{\boldsymbol{\alpha}})]^T \tag{9}$$

The cost function consists of two parts: the first part evaluates the error of the fitting, and the second part evaluates the sparsity of the parameter vector $\bar{\boldsymbol{\alpha}}$. Then we can estimate the parameter vector $\boldsymbol{\alpha}$ by solving the sparsity-regularized nonlinear least squares problem

$$\min_{\bar{\boldsymbol{\alpha}}} \mathcal{L}_\lambda(\mathbf{x}_0, \bar{\boldsymbol{\alpha}}) \tag{10}$$

Due to the nonlinearity in the chaotic system, the cost function $\mathcal{L}_\lambda(\mathbf{x}_0, \bar{\boldsymbol{\alpha}})$ may contain local minima. In the estimation of dynamical system parameters, the multiple shooting (MS) method [Peifer & Timmer, 2007] is effective to reduce the problem of local minima. Considering the signal sparsity, we combine the MS method with the iteratively reweighted nonlinear least squares (IRNLS) in [Liu *et al.*, 2012] and develop a MS-IRNLS algorithm for solving (10). An algorithmic framework is given in Appendix A.

# IV. Analysis on Reconstructability with the Parameter Identifiability

In this section, we conduct the analysis on reconstructability of the ChaA2I from identifiability of system parameters.

**A. Parameter Identifiability**

The parameter identifiability is defined as the ability to uniquely determine the system parameters from a given set of measurement data [Carson, *et al.*, 1983]. For our problem, we have the following three cases according to the solutions of the parameter estimation problem [Audoly *et al.*, 2001]: (1) *Globally* identifiable if and only if the parameter estimation problem has the only solution; (2) *Locally* identifiable if and only if the parameter estimation problem has finite number of solutions; and (3) *Non-identifiable* if and only if the parameter estimation problem has infinite number of solutions.

For the parameter estimation of the nonlinear dynamic system, it is quite difficult to determine whether the parameters are globally identifiable or not. Even the parameter estimation problem is globally identifiable, the global optimal solution is almost impossible to acquire. For these reasons, we focus on the local identifiability of the parameter estimation problem.

The local identifiability is traditionally concerned with nonlinear least squares model[1]

$$\min_{\bar{\boldsymbol{\alpha}}} \mathcal{L}_0(\mathbf{x}_0, \bar{\boldsymbol{\alpha}}) = \|\mathbf{y} - \mathcal{H}(\mathbf{x}_0, \bar{\boldsymbol{\alpha}})\|_2^2 \qquad (11)$$

Let $B_\rho(\boldsymbol{\alpha}) = \{\bar{\boldsymbol{\alpha}} \|\bar{\boldsymbol{\alpha}} - \boldsymbol{\alpha}\|_2 < \rho\}$ be the neighborhood of the parameter vector $\boldsymbol{\alpha}$ with radius $\rho$. The parameter vector $\boldsymbol{\alpha}$ is locally identifiable if (11) has an unique solution among the neighborhood $B_\rho(\boldsymbol{\alpha})$.

The numerical method for checking the local identifiability is, firstly introduced in [Jacquez & Greif, 1985], based on the linearization of $\mathcal{H}(\mathbf{x}_0, \bar{\boldsymbol{\alpha}})$ around the actual value $\boldsymbol{\alpha}$ and the analysis of its sensitivity matrix $\mathcal{S}(\mathbf{x}_0, \boldsymbol{\alpha})$

$$\mathcal{S}(\mathbf{x}_0, \boldsymbol{\alpha}) \triangleq \left. \frac{\partial \mathcal{H}(\mathbf{x}_0, \bar{\boldsymbol{\alpha}})}{\partial \bar{\boldsymbol{\alpha}}^T} \right|_{\boldsymbol{\alpha}} \qquad (12)$$

With the linearization of $\mathcal{H}(\mathbf{x}_0, \bar{\boldsymbol{\alpha}})$ around $\boldsymbol{\alpha}$,

$$\mathcal{H}(\mathbf{x}_0, \bar{\boldsymbol{\alpha}}) \approx \mathcal{H}(\mathbf{x}_0, \boldsymbol{\alpha}) + \mathcal{S}(\mathbf{x}_0, \boldsymbol{\alpha})(\bar{\boldsymbol{\alpha}} - \boldsymbol{\alpha}) \qquad (13)$$

the nonlinear least squares problem (11) is locally approximated as a linear least squares one

$$\min_{\bar{\boldsymbol{\alpha}} \in B_\rho(\boldsymbol{\alpha})} \|\mathbf{y} - \mathcal{H}(\mathbf{x}_0, \boldsymbol{\alpha}) - \mathcal{S}(\mathbf{x}_0, \boldsymbol{\alpha})(\bar{\boldsymbol{\alpha}} - \boldsymbol{\alpha})\|_2^2 \qquad (14)$$

Then (14) has an unique solution if and only if the columns of the sensitivity matrix $\mathcal{S}(\mathbf{x}_0, \boldsymbol{\alpha})$ are linearly independent. Denote $\mathcal{G}(\mathbf{x}_0, \boldsymbol{\alpha})$ as the correlation matrix [Rodgers et al., 1988] with element $g_{ij}(\mathbf{x}_0, \boldsymbol{\alpha})$ ($1 \leq i, j \leq B$) by

$$g_{ij}(\mathbf{x}_0, \boldsymbol{\alpha}) = \frac{\langle (\mathbf{s}_i(\mathbf{x}_0, \boldsymbol{\alpha}) - \bar{s}_i(\mathbf{x}_0, \boldsymbol{\alpha})\mathbf{1}_M), (\mathbf{s}_j(\mathbf{x}_0, \boldsymbol{\alpha}) - \bar{s}_j(\mathbf{x}_0, \boldsymbol{\alpha})\mathbf{1}_M) \rangle}{\|\mathbf{s}_i(\mathbf{x}_0, \boldsymbol{\alpha}) - \bar{s}_i(\mathbf{x}_0, \boldsymbol{\alpha})\mathbf{1}_M\|_2 \|\mathbf{s}_j(\mathbf{x}_0, \boldsymbol{\alpha}) - \bar{s}_j(\mathbf{x}_0, \boldsymbol{\alpha})\mathbf{1}_M\|_2} \qquad (15)$$

where $\mathbf{s}_i(\mathbf{x}_0, \boldsymbol{\alpha})$ is the $i$th column of $\mathcal{S}(\mathbf{x}_0, \boldsymbol{\alpha})$, $\bar{s}_i(\mathbf{x}_0, \boldsymbol{\alpha})$ is the mean value of the elements in $\mathbf{s}_i(\mathbf{x}_0, \boldsymbol{\alpha})$, $\mathbf{1}_M = [1,1,\ldots,1]^T \in \mathbb{R}^M$. Let $\mu_{\mathcal{S}}(\mathbf{x}_0, \boldsymbol{\alpha})$ be the maximum absolute value of the off-diagonal elements of the correlation matrix $\mathcal{G}(\mathbf{x}_0, \boldsymbol{\alpha})$

$$\mu_{\mathcal{S}}(\mathbf{x}_0, \boldsymbol{\alpha}) = \max_{i \neq j} \left| g_{ij}(\mathbf{x}_0, \boldsymbol{\alpha}) \right| \tag{16}$$

---

[1] In traditional analysis of identifiability, it is assumed that the measurement is overdetermined, *i.e.* the number of measurements are greater than that of the variables. In our case, the problem is underdetermined. To avoid the introduction of new variables, we assume that the measurement is overdetermined in this subsection.

It has been shown [Zak et al., 2003] that the columns of the sensitivity matrix $\mathcal{S}(\mathbf{x}_0, \boldsymbol{\alpha})$ are linearly independent if $\mu_{\mathcal{S}}(\mathbf{x}_0, \boldsymbol{\alpha}) < 1$. That is that the parameter vector $\boldsymbol{\alpha}$ is locally identifiable among the neighborhood $B_\rho(\boldsymbol{\alpha})$ if $\mu_{\mathcal{S}}(\mathbf{x}_0, \boldsymbol{\alpha}) < 1$. However, if $\mu_{\mathcal{S}}(\mathbf{x}_0, \boldsymbol{\alpha})$ is close to 1, the sensitivity matrix $\mathcal{S}(\mathbf{x}_0, \boldsymbol{\alpha})$ is ill-conditioned and (14) may lead to a poor estimation of $\boldsymbol{\alpha}$. Thus, it is expected that we can obtain a good estimation of $\boldsymbol{\alpha}$ for a small $\mu_{\mathcal{S}}(\mathbf{x}_0, \boldsymbol{\alpha})$.

**B. Analysis on Reconstructability**

The parameter identifiability in the last subsection cannot be directly used to study the reconstructability of the ChaA2I converter. A difficulty is that the reconstruction cost (8) is not differentiable. To circumvent the problem, we can approximate the $l_0$-norm with weighted $l_2$-norm as in the development of $l_0$ regularized algorithms. For $\bar{\boldsymbol{\alpha}} \in B_\rho(\boldsymbol{\alpha})$, $\|\bar{\boldsymbol{\alpha}}\|_0$ can be approximated as [Candès et al., 2008]

$$\|\bar{\boldsymbol{\alpha}}\|_0 \approx \left\| \mathbf{W}^{1/2} \bar{\boldsymbol{\alpha}} \right\|_2^2 \tag{17}$$

where $\mathbf{W} = diag([w_1, w_2, \ldots, w_B])$ is the weighted matrix with

$$w_i = \frac{1}{\alpha_i^2 + \varepsilon} \quad (1 \leq i \leq B) \tag{18}$$

In (18), $\varepsilon > 0$ is a small positive constant such that the weighted matrix $\mathbf{W}$ is positive definite. With (17), finding the sparse solution of (8) is equivalent to solving

$$\min_{\bar{\boldsymbol{\alpha}} \in B_\rho(\boldsymbol{\alpha})} \left\| \begin{bmatrix} \mathbf{y} \\ 0 \end{bmatrix} - \begin{bmatrix} \mathcal{H}(\mathbf{x}_0, \bar{\boldsymbol{\alpha}}) \\ \lambda^{1/2} \mathbf{W}^{1/2} \bar{\boldsymbol{\alpha}} \end{bmatrix} \right\|_2^2 \quad (19)$$

Because of the regularization approximation, (19) is an overdetermined problem. With the definition (12) and the linearization of $\mathcal{H}(\mathbf{x}_0, \bar{\boldsymbol{\alpha}})$ around $\boldsymbol{\alpha}$ in (13), we have the sparsity-regularized sensitivity matrix of (19) as

$$\mathcal{S}^\lambda(\mathbf{x}_0, \boldsymbol{\alpha}) = \begin{bmatrix} \mathcal{S}(\mathbf{x}_0, \boldsymbol{\alpha}) \\ \lambda^{1/2} \mathbf{W}^{1/2} \end{bmatrix} \quad (20)$$

Then we can obtain the correlation matrix $\mathcal{G}^\lambda(\mathbf{x}_0, \boldsymbol{\alpha})$ of $\mathcal{S}^\lambda(\mathbf{x}_0, \boldsymbol{\alpha})$ with the element $g_{ij}^\lambda(\mathbf{x}_0, \boldsymbol{\alpha})$ ($1 \leq i, j \leq B$) by

$$\begin{aligned} g_{ij}^\lambda(\mathbf{x}_0, \boldsymbol{\alpha}) &= \frac{\left\langle \left(\mathbf{s}_i^\lambda(\mathbf{x}_0, \boldsymbol{\alpha}) - \bar{s}_i^\lambda(\mathbf{x}_0, \boldsymbol{\alpha})\mathbf{1}_M\right), \left(\mathbf{s}_j^\lambda(\mathbf{x}_0, \boldsymbol{\alpha}) - \bar{s}_j^\lambda(\mathbf{x}_0, \boldsymbol{\alpha})\mathbf{1}_M\right) \right\rangle}{\left\| \mathbf{s}_i^\lambda(\mathbf{x}_0, \boldsymbol{\alpha}) - \bar{s}_i^\lambda(\mathbf{x}_0, \boldsymbol{\alpha})\mathbf{1}_M \right\|_2 \left\| \mathbf{s}_j^\lambda(\mathbf{x}_0, \boldsymbol{\alpha}) - \bar{s}_j^\lambda(\mathbf{x}_0, \boldsymbol{\alpha})\mathbf{1}_M \right\|_2} \\ &= \frac{\left\langle \mathbf{s}_i(\mathbf{x}_0, \boldsymbol{\alpha}), \mathbf{s}_j(\mathbf{x}_0, \boldsymbol{\alpha}) \right\rangle - N \bar{s}_i^\lambda(\mathbf{x}_0, \boldsymbol{\alpha}) \bar{s}_j^\lambda(\mathbf{x}_0, \boldsymbol{\alpha})}{\sqrt{\left\langle \mathbf{s}_i(\mathbf{x}_0, \boldsymbol{\alpha}), \mathbf{s}_i(\mathbf{x}_0, \boldsymbol{\alpha}) \right\rangle + \lambda w_i - N\left(\bar{s}_i^\lambda(\mathbf{x}_0, \boldsymbol{\alpha})\right)^2}} \times \\ &\quad \frac{1}{\sqrt{\left\langle \mathbf{s}_j(\mathbf{x}_0, \boldsymbol{\alpha}), \mathbf{s}_j(\mathbf{x}_0, \boldsymbol{\alpha}) \right\rangle + \lambda w_j - N\left(\bar{s}_j^\lambda(\mathbf{x}_0, \boldsymbol{\alpha})\right)^2}} \end{aligned} \quad (21)$$

where $\bar{s}_i^\lambda(\mathbf{x}_0, \boldsymbol{\alpha})$ is the mean value of the elements in $\mathbf{s}_i^\lambda(\mathbf{x}_0, \boldsymbol{\alpha})$ and $N = M + B$. Let $\mu_{\mathcal{S}^\lambda}(\mathbf{x}_0, \boldsymbol{\alpha})$ be the maximum absolute value of the off-diagonal elements of the correlation matrix $\mathcal{G}^\lambda(\mathbf{x}_0, \boldsymbol{\alpha})$

$$\mu_{\mathcal{S}^\lambda}(\mathbf{x}_0, \boldsymbol{\alpha}) = \max_{i \neq j} \left| g_{ij}^\lambda(\mathbf{x}_0, \boldsymbol{\alpha}) \right| \quad (22)$$

With the discussions in last subsection, we have the following reconstructable condition.

**Theorem:** The sparse signal is locally reconstructable by solving the sparsity-regularized nonlinear least squares problem (10) if $\mu_{\mathcal{S}^\lambda}(\mathbf{x}_0, \boldsymbol{\alpha}) < 1$. The condition is called as correlation-based reconstructable condition (CRC).

*Proof:* See Appendix B.

In practice, we do not know the initial state $\mathbf{x}_0$. A possible way is to compute the average of $\bar{\mu}_{\mathcal{S}^\lambda}(\boldsymbol{\alpha})$ over the state space as

$$\bar{\mu}_{\mathcal{S}^\lambda}(\boldsymbol{\alpha}) = \frac{1}{N_{\exp}} \sum_{\mathbf{x}_0 \in \mathcal{X}_0} \mu_{\mathcal{S}^\lambda}(\mathbf{x}_0, \boldsymbol{\alpha}) \tag{23}$$

where $\mathcal{X}_0$ is the set of the initial states in the experiments and $N_{\exp} = |\mathcal{X}_0|$ denotes the times of the experiments.

As revealed in (20), the CRC depends on the chaotic system, the sparse signal and the regularization parameter $\lambda$. Since the sampling interval $T_{cs}$ is implicitly contained in the measurement $\mathcal{H}(\mathbf{x}_0, \boldsymbol{\alpha})$, the CRC will also depend on the sampling interval $T_{cs}$. Then the appropriate sampling interval can be determined from the CRC by studying local reconstructability of the sparsity-regularized nonlinear least squares problem (10). For a given chaotic system, the sampling interval, the sparse level and the parameter $\lambda$ will determine the identifiability. In Section V, we will simulate their effects on the reconstructability.

### C. Remarks on the Reconstructability

At the first glance, the CRC seems to be the coherence [Donoho & Elad, 2003, Davenport *et al.*, 2012] of the measurement matrix in the compressive sensing theory with the linear measurements. Both of them compute the similarities between any two columns of a matrix. However, the matrices in the two theories are quite different. In linear CS, the coherence is defined as the largest absolute inner product between any two columns $\mathbf{M}_i$, $\mathbf{M}_j$ of the measurement matrix $\mathbf{M}$

$$c_{\mathbf{M}} = \max_{i \neq j} \frac{|\langle \mathbf{M}_i, \mathbf{M}_j \rangle|}{\|\mathbf{M}_i\|_2 \|\mathbf{M}_j\|_2} \tag{24}$$

The coherence $c_\mathbf{M}$ indicates that the linear independent columns of $\mathbf{M}$ are no more than $1+c_\mathbf{M}^{-1}$ [Donoho & Elad, 2003]. For the reconstruction of the $M$-sparse signal, it is required that $M < (1+c_\mathbf{M}^{-1})/2$ to keep any $2M$ columns of the measurement matrix $\mathbf{M}$ to be linear independent [Davenport et al., 2012].

The CRC condition is derived from the least squares formulation. The $\mu_{\mathcal{S}^\lambda}(\mathbf{x}_0, \boldsymbol{\alpha})$ measures the similarities between the columns of $\mathcal{S}^\lambda(\mathbf{x}_0, \boldsymbol{\alpha})$. With $\mu_{\mathcal{S}^\lambda}(\mathbf{x}_0, \boldsymbol{\alpha}) < 1$, the matrix $\mathcal{S}^\lambda(\mathbf{x}_0, \boldsymbol{\alpha})$ is of full column rank and the least squares problem (19) has unique solution. In this sense, the CRC does not directly refer to the reconstructability of the linearized measurement and is restricted to the formulation (19).

In the development of the linear CS, the CS theory with nonlinear measurements also receives attention [Blumensath, 2010; Bahmani et al., 2012; Blumensath, 2012]. Parallel to the RIP conditions, some RIP-like reconstruction conditions are derived. Different from the linear CS, the nonlinear reconstructions are conducted from iterative algorithms and the measurement matrix must be updated with step-by-step. To be reconstructable, it is required that the linearized measurement matrix satisfies the RIP-like conditions at each step. For practical applications, the confirmation of the conditions is troublesome and no real examples are reported. The chaotic CS utilizes the chaotic systems to implement the nonlinear measurement. To the best of our knowledge, it is the first nonlinear CS to be implemented.

## V. Simulation Results

In this section, we take Lorenz system as an example to study the performance of the proposed ChaA2I converter. The effects of the signal sparsity, the regularization parameter and the sampling interval on the reconstructability are evaluated. The reconstruction performance is demonstrated.

**A. Simulation Setting**

The frequency-sparse signal $s(t)$ is assumed to have the bandwidth $50Hz$ with $B=100$. Then the Fourier coefficient vector is $\boldsymbol{\alpha} \in \mathbb{R}^{100}$. To generate the $W$-sparse $\boldsymbol{\alpha}$, the positions of $W$ nonzero elements in $\boldsymbol{\alpha}$ are uniformly distributed among the 100 dimensions of $\boldsymbol{\alpha}$ and the amplitudes of the $W$ nonzero elements are from two types of distributions (*i.i.d.* Gaussian with zero mean and unity variance and *i.i.d.* Bernoulli with entries $\pm 1$). For the sparse signal, the Nyquist sampling interval $T_{nq}$ is equal to $0.01s$.

The Lorenz system excited by the signal $s(t)$ is represented as

$$\begin{cases} \dot{x}_1(t) = \tau\left(a\left(x_2(t)-x_1(t)\right)\right) \\ \dot{x}_2(t) = \tau\left(bx_1(t)-x_2(t)-x_1(t)x_3(t)\right)+\mu s(t) \\ \dot{x}_3(t) = \tau\left(x_1(t)x_2(t)-cx_3(t)\right) \end{cases} \quad (25)$$

where $s(t)$ is added to the state $x_2$ through the coupling strength $\mu$. $a$, $b$, and $c$ are the system parameters. It is assumed that the state $x_2$ is observable and is used to perform the compressive measurements. In the simulation, $(a,b,c)=(10,28,2.66)$ and $\mu=20$. For the parameter setting, the Lorenz system works in chaotic state. In (25), $\tau$ is a time-scaling factor which controls the changing rate of the chaotic states. The larger the factor $\tau$ is, the larger the system bandwidth will be. Fig.2 shows the relation between the time-scaling factor $\tau$ and the bandwidth[2] of the Lorenz system. Intuitively, the bandwidth should be at least the bandwidth of the input signal so that the output encompasses the input information. With the simulation parameters, we choose $\tau=15$, with which the bandwidth of the observable output $x_2$ is $51Hz$.

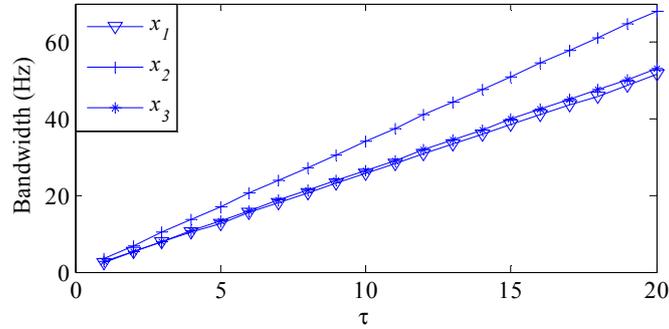

Fig.2 Bandwidth of Lorenz system.

**B. Recontructability**

This subsection firstly simulates the effects of the signal sparsity and the regularization parameter on the reconstructability, and then provides a guideline for the determination of the sampling interval $T_{cs}$.

**a) Dependence of $\bar{\mu}_{\mathcal{S}^\lambda}(\boldsymbol{\alpha})$ on Signal Sparsity**

In the first experiment, we fix the regularization parameter $\lambda$ ($=10^{-6}$) and compute the distributions of $\bar{\mu}_{\mathcal{S}^\lambda}(\boldsymbol{\alpha})$ with different sparsity (positions, amplitudes and levels) under different sampling intervals. $1\times10^3$ independent experiments are conducted. The average results are shown in Fig.3 and Fig.4. It is seen that $\bar{\mu}_{\mathcal{S}^{\lambda}}(\boldsymbol{\alpha})$ with the different sparsity has almost the same distribution under the same sampling interval. The distributions of $\bar{\mu}_{\mathcal{S}^{\lambda}}(\boldsymbol{\alpha})$ are closely related to the chaotic system and the sampling intervals, but are almost not affected by the signal sparsity. In this case ($\lambda = 10^{-6}$), the regulation term $\lambda \|\boldsymbol{\alpha}\|_0$ has less effect on the value of $\bar{\mu}_{\mathcal{S}^{\lambda}}(\boldsymbol{\alpha})$ and the reconstructability by the sub-Nyquist samples is poor because of large $\bar{\mu}_{\mathcal{S}^{\lambda}}(\boldsymbol{\alpha})$.

---

[2] The bandwidth of chaotic signal is defined as the spectrum bandwidth that concentrates 99% energy of the signal [Willsey et al., 2011].

For large $\lambda$, the weighted matrix $\mathbf{W}$ (18) affects the value of $\bar{\mu}_{\mathcal{S}^{\lambda}}(\boldsymbol{\alpha})$. The results are shown in Fig.5 and Fig.6 for $\lambda = 10^{-2}$. In this case, $\bar{\mu}_{\mathcal{S}^{\lambda}}(\boldsymbol{\alpha})$ have different distributions for different sparsity levels. The effects of the regulation term $\lambda \|\boldsymbol{\alpha}\|_0$ on the value of $\bar{\mu}_{\mathcal{S}^{\lambda}}(\boldsymbol{\alpha})$ are obvious. In comparison with the results in Fig. 4 and Fig.6, we may find that value of $\bar{\mu}_{\mathcal{S}^{\lambda}}(\boldsymbol{\alpha})$ decreases as the sparsity level decreases. It means that the reconstructability is better for lower sparsity level $W$.

It is also noted that the distribution ranges of $\bar{\mu}_{\mathcal{S}^{\lambda}}(\boldsymbol{\alpha})$ increase as the sampling interval increases. According to the analysis of the reconstructability, we can conclude that reconstructability of the ChaA2I converter becomes more difficult as the sampling interval increases.

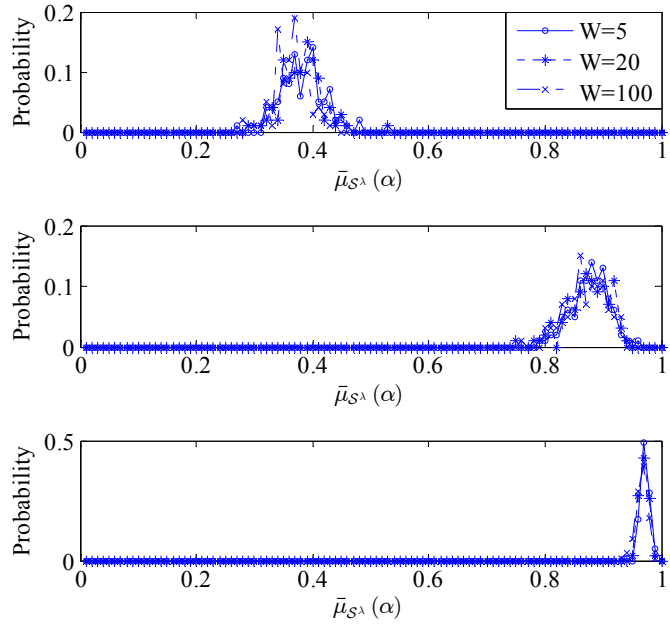

Fig.3 The distribution of $\bar{\mu}_{\mathcal{S}^\lambda}(\boldsymbol{\alpha})$ with $\lambda = 10^{-6}$ for Gaussian $\boldsymbol{\alpha}$. Upper: $T_{cs} = T_{nq} = 0.01s$; Middle: $T_{cs} = 0.02s$; Lower: $T_{cs} = 0.04s$.

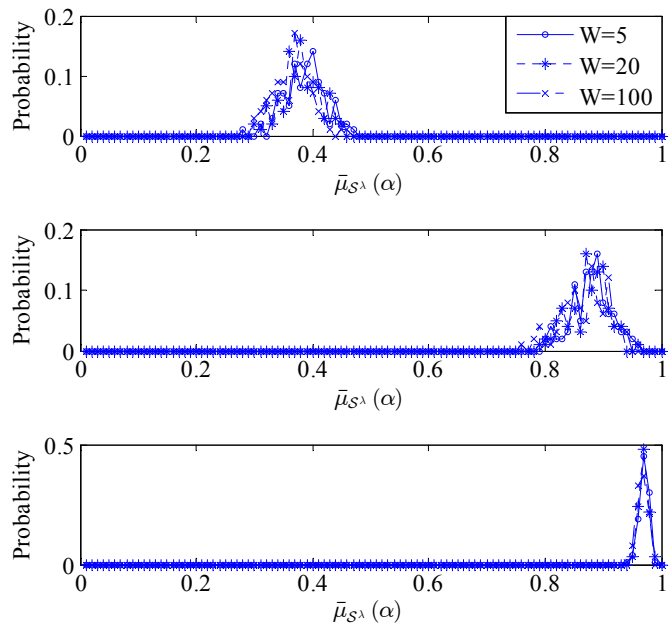

Fig.4 The distribution of $\bar{\mu}_{\mathcal{S}^\lambda}(\boldsymbol{\alpha})$ with $\lambda = 10^{-6}$ for Bernoulli $\boldsymbol{\alpha}$. Upper: $T_{cs} = T_{nq} = 0.01s$; Middle: $T_{cs} = 0.02s$; Lower: $T_{cs} = 0.04s$.

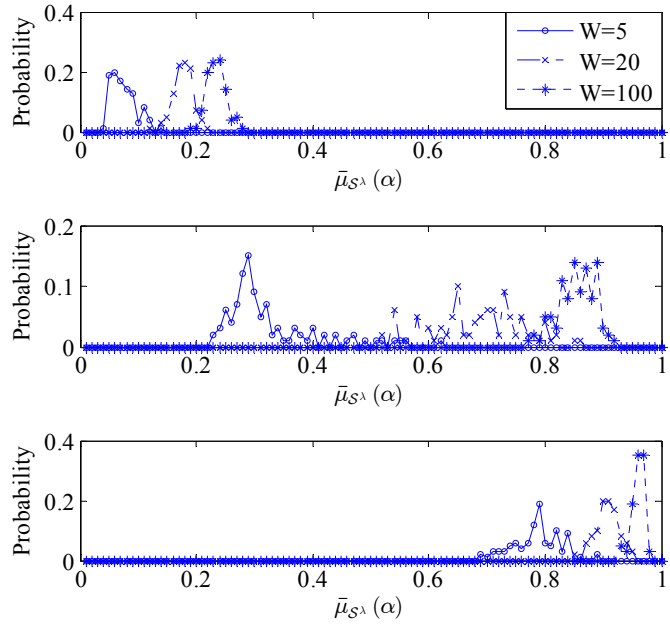

Fig.5 The distribution of $\bar{\mu}_{\mathcal{S}^\lambda}(\mathbf{\alpha})$ with $\lambda = 10^{-2}$ for Gaussian $\mathbf{\alpha}$.
Upper: $T_{cs} = T_{nq} = 0.01s$; Middle: $T_{cs} = 0.02s$; Lower: $T_{cs} = 0.04s$.

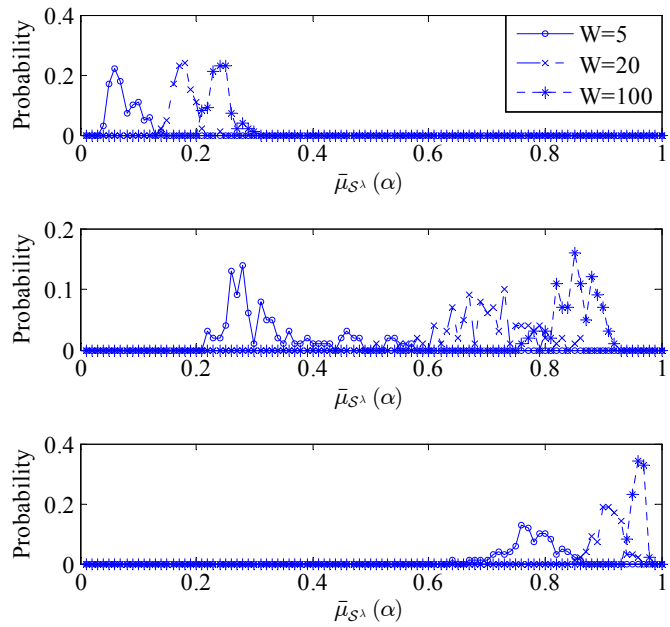

Fig.6 The distribution of $\bar{\mu}_{\mathcal{S}^\lambda}(\mathbf{\alpha})$ with $\lambda = 10^{-2}$ for Bernoulli $\mathbf{\alpha}$.
Upper: $T_{cs} = T_{nq} = 0.01s$; Middle: $T_{cs} = 0.02s$; Lower: $T_{cs} = 0.04s$.

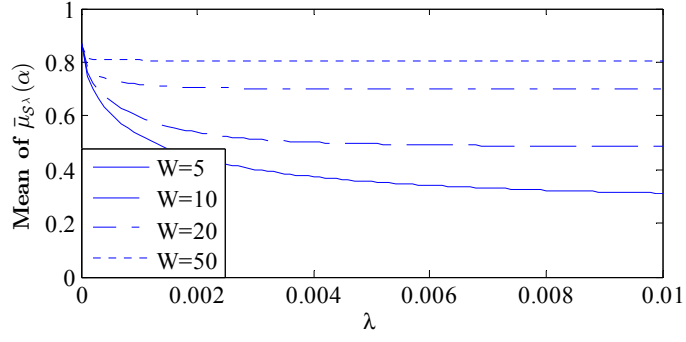

(a)

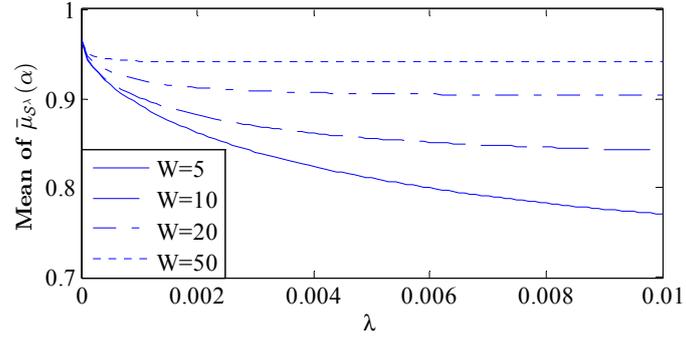

(b)

Fig.7 The mean of $\bar{\mu}_{\mathcal{S}^\lambda}(\boldsymbol{\alpha})$ with the different regularization parameter $\lambda$.

(a) $T_s = 0.02s$ and (b) $T_s = 0.04s$

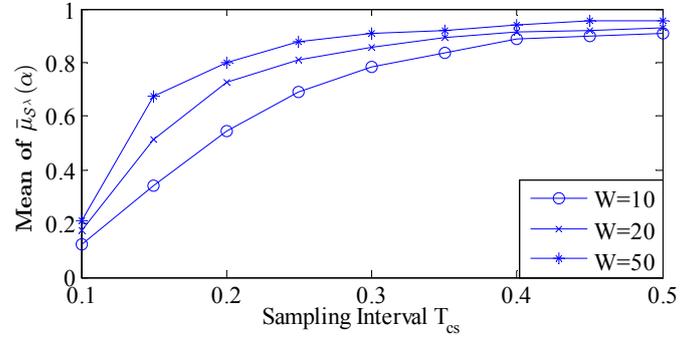

Fig.8 The mean of $\bar{\mu}_{\mathcal{S}^\lambda}(\boldsymbol{\alpha})$ with the different sampling intervals.

**b) Dependence of $\bar{\mu}_{\mathcal{S}_\lambda}(\boldsymbol{\alpha})$ on Regularization Parameter**

As noted in (20) and simulations in last subsection, the larger the $\lambda$ is, the smaller the $\bar{\mu}_{\mathcal{S}^\lambda}(\boldsymbol{\alpha})$ is. Fig. 7 shows the variation of $\bar{\mu}_{\mathcal{S}^\lambda}(\boldsymbol{\alpha})$ as the regularization parameter $\lambda$ for different sparsity levels $W$. It can be seen that for a given sparse level, the $\bar{\mu}_{\mathcal{S}^\lambda}(\boldsymbol{\alpha})$ decreases much fast as the $\lambda$ increases from zero and gradually converge. It means that we can select large $\lambda$ to improve the reconstructability. However, in the family of regularized least squares algorithms [Wainwright, 2009], $\lambda$ plays the role in balancing the sparse levels and least squares fitting. The appropriate $\lambda$ should be the small positive value that keeps the value of $\bar{\mu}_{\mathcal{S}^\lambda}(\boldsymbol{\alpha})$ as low as possible. Therefore, there is a trade-off for the selection of $\lambda$. For the simulated system, we may select $\lambda = 2\times 10^{-3}$ with the unknown sparse level. Fig. 7 again reveals that the reconstructability becomes difficult for large sparse level $W$.

**c) Determination of $T_{cs}$ from the Reconstructability**

With the analyses of Parts a) and b) on the reconstructability, we simulate the determination of the sampling interval $T_{cs}$. Fig. 8 shows the variation of $\bar{\mu}_{\mathcal{S}^\lambda}(\boldsymbol{\alpha})$ as the sampling interval $T_{cs}$ for different sparsity levels $W$ with the regularization parameter as $\lambda = 2\times 10^{-3}$. It is seen that as the sampling interval $T_{cs}$ increases, the mean of $\bar{\mu}_{\mathcal{S}^\lambda}(\boldsymbol{\alpha})$ increases and asymptotically approaches 1. It is also noted that the mean of $\bar{\mu}_{\mathcal{S}^\lambda}(\boldsymbol{\alpha})$ increases much slowly when the sampling interval $T_{cs}$ is larger than some thresholds. This is because the measurement data $y_m$ do not contain enough information of the excitation signal for the larger $T_{cs}$, and can not ensure the reconstructability. With these observations, we can select the sampling interval to be no larger than the threshold value. For the simulated example, $T_{cs}$ can be set to be no more than 0.04s and 0.035s for $W = 10$ and $W = 20$, respectively. In the next subsection, we will evaluate the reconstruction performance of the ChaA2I system for $T_{cs} = 0.02s$, $0.03s$ and $0.04s$.

## C. Reconstruction Performance

In this subsection, we evaluate the sparse signal reconstruction performance of the ChaA2I. The MS-IRNLS algorithm in Appendix A is used to compute the sparse coefficients. Each element of the initial searching points $\tilde{\boldsymbol{\alpha}}$ is randomly chosen over interval [-1, 1]. The initial state of the Lorenz system is randomly chosen over the range of the attractor of the Lorenz system. $\lambda$ and $\varepsilon$ are set to be $2\times 10^{-3}$ and $10^{-3}$, respectively. The MS-IRNLS is deemed convergent if the absolute error (stopping criterion) between two consecutive iterations is less than $10^{-3}$. The solution by the MS-IRNLS with one set of initial settings is called one realization. The relative error defined by $Err = \|\hat{\boldsymbol{\alpha}} - \boldsymbol{\alpha}\|_2^2 / \|\boldsymbol{\alpha}\|_2^2$ is used to measure reconstruction performance of the sparse signals, where $\hat{\boldsymbol{\alpha}}$ is the estimated $\boldsymbol{\alpha}$ through the MS-IRNLS algorithm. To reduce the effect of local minima, in each experiments for a given $\boldsymbol{\alpha}$, 20 realizations are performed with the different initial searching points and the different initial states. The minimum relative error among the set of the estimated $\boldsymbol{\alpha}$ is selected to evaluate the reconstruction performance. The results are averaged over 100 experiments for the different values of the Fourier representation vector $\boldsymbol{\alpha}$.

Figures 9 and 10 show one realization for the case of the Gaussian coefficients with $T_{cs} = 0.02s$. Two sparsity levels ($W = 5$ and $W = 15$) are simulated. Sparsity positions and amplitudes are shown in Fig.9 (a) and Fig.10 (a), respectively. It is seen from Fig.9 that for $W = 5$, the reconstructed signal matches the original one well and sparse positions/coefficients are estimated correctly. The estimated relative error is $1.4\times 10^{-3}$. While for $W = 15$, the estimated relative error increases to $2.9\times 10^{-3}$. In both cases, the reconstructed signals are reasonable approximations to the original one (Fig.9(b) and Fig.10(b)). From the point of view of waveform reconstruction, the sparse signal is well reconstructed, although the sampling rate is only $1/2$ of the Nyquist rate. Fig.11 and Fig.12 show the results similar to those in Fig.9 and Fig.10 for the Bernoulli coefficients.

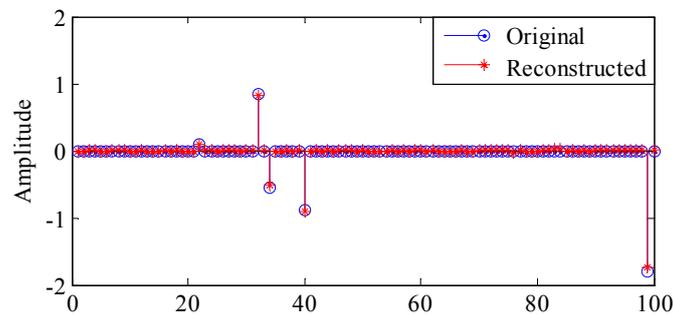

(a)

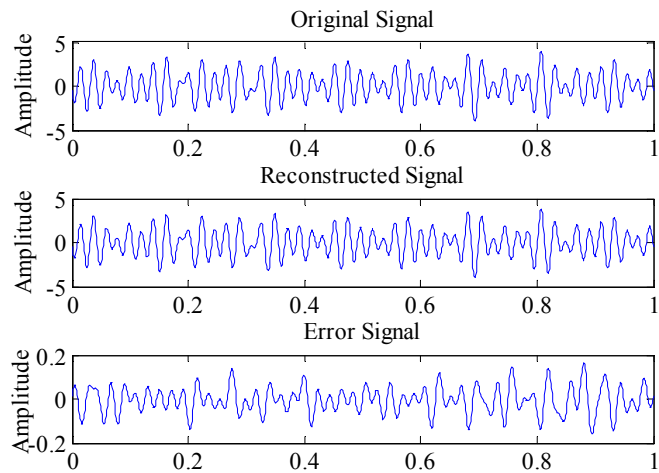

(b)

Fig.9 Sparse signal and its reconstruction for $W = 5$ with Gaussian distribution. (a) Fourier coefficient vector and its estimation and (b) signal waveform and its estimation.

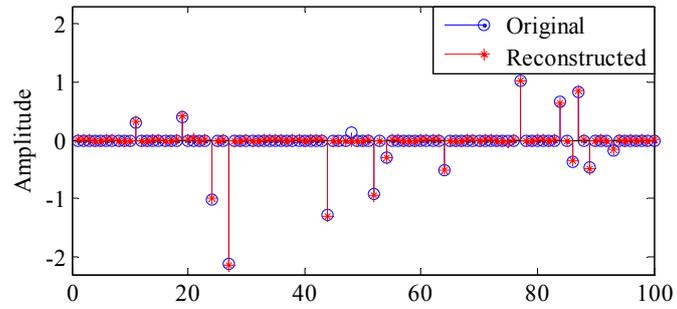

(a)

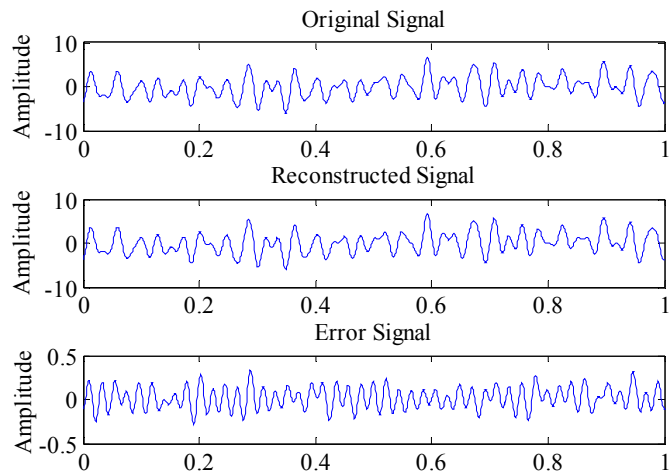

(b)

Fig.10 Sparse signal and its reconstruction for $W=15$ with Gaussian distribution. (a) Fourier coefficient vector and its estimation and (b) signal waveform and its estimation.

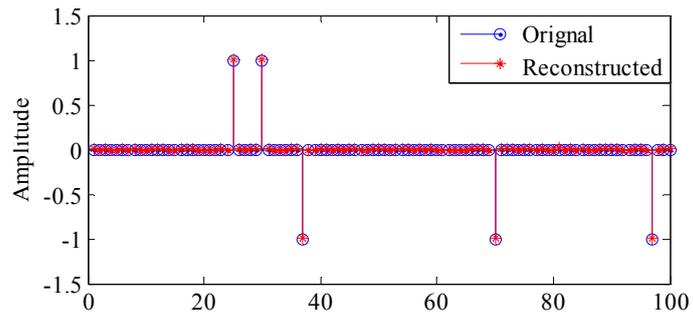

(a)

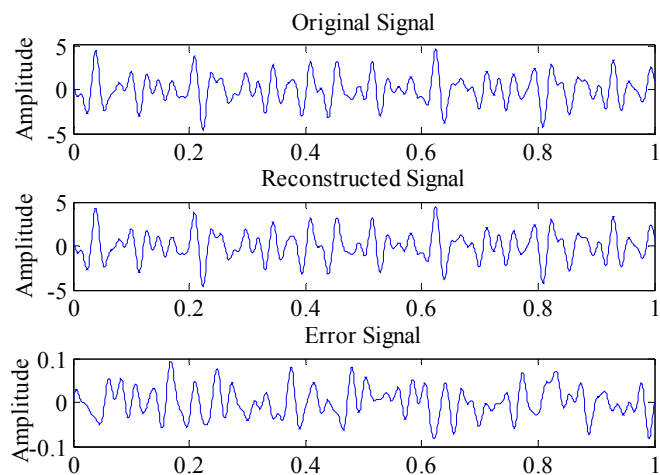

(b)

Fig.11 Sparse signal and its reconstruction for $W = 5$ with Bernoulli distribution. (a) Fourier coefficient vector and its estimation and (b) signal waveform and its estimation

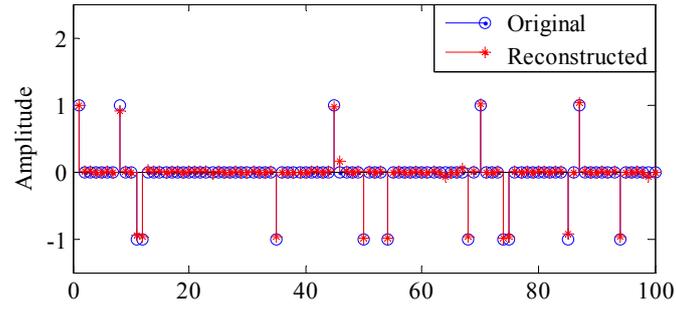

(a)

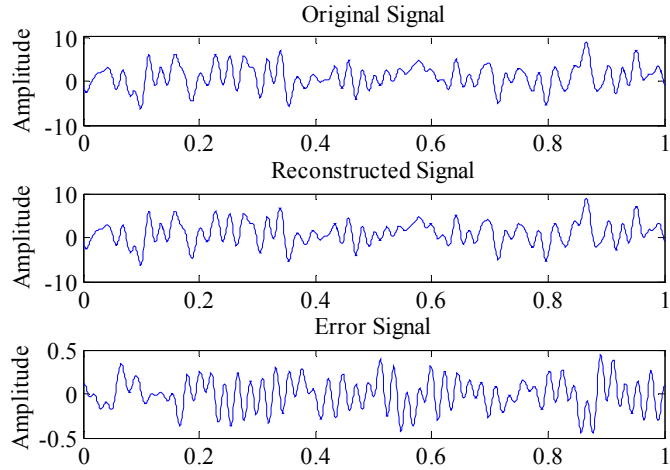

(b)

Fig.12 Sparse signal and its reconstruction for $W=15$ with Bernoulli distribution. (a) Fourier coefficient vector and its estimation and (b) signal waveform and its estimation

In Fig.13, we show the relative error of the sparse signal reconstruction with the different sparsity levels $W$ and the different distributions of $\boldsymbol{\alpha}$. For small sparsity level $W$, the reconstruction error keeps low even with the sampling interval $T_{cs}=0.04s$ (1/4 Nyquist rate). However, the reconstruction performance gets poor when the sparsity level $W$ is large. These results are consistent with the analysis of reconstructability in last Section.

Also seen from Fig.13, the reconstruction performance of the Bernoulli distribution is superior to that of the Gaussian one for small sparsity level $W$. However, for large sparsity level $W$, the reconstruction performance of the Gaussian distribution is superior. This is an inherent phenomenon in the sparsity-regularized algorithms. For the Bernoulli distribution, all of the nonzero elements of $\boldsymbol{\alpha}$ have the same amplitudes. Thus the $\boldsymbol{\alpha}$ can be estimated accurately once the nonzero positions is identified. For the Gaussian distribution, it is required to estimate not only the nonzero positions but also the amplitudes of the nonzero elements of $\boldsymbol{\alpha}$.

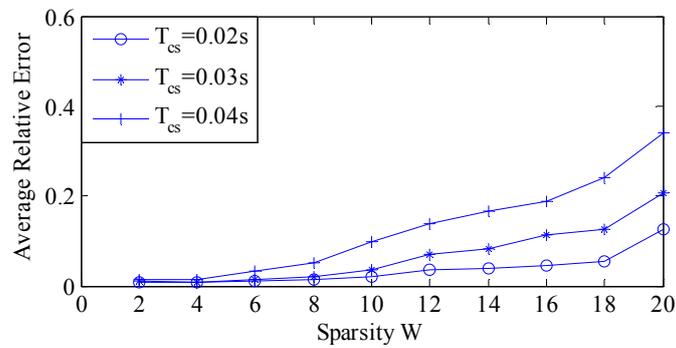

(a)

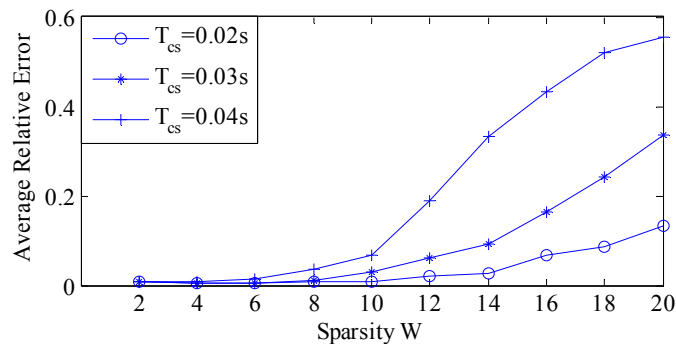

(b)

Fig.13 Average relative errors vs. sparsity of the proposed ChaA2I converter with different sampling intervals. (a) Gaussian distribution and (b) Bernoulli distribution.

# VI. Conclusions

We presented the first chaos-based A2I conversion for the acquisition and reconstruction of sparse analog signals. The conversion principle, reconstructability and reconstruction performance are studied. Different from the linear CS, the randomness of the sparse signals is through chaotic systems. Along with the deterministic randomness, the reconstructability is analyzed from the point of view of identifiability of system parameters. It is shown that the sparse signal is locally reconstructable if the columns of the sparsity-regularized sensitivity matrix are linearly independent. The extensive simulations prove the rationality of the analysis on reconstructability and demonstrate the effectiveness of the ChaA2I to implement the sub-Nyquist sampling of the analog signals.

A practical problem with the ChaA2I is the development of the reconstruction algorithms. Because of the nonlinearity in the formulation, there are no global algorithms for the effective reconstruction. Developing the effective sparse reconstruction algorithms is the aim of our future works.

## Appendix A: MS-IRNLS Algorithm

In this appendix, we combine the MS method with the IRNLS algorithm [Liu *et al.*, 2012] to solve (10). The basic idea of MS-IRNLS is to divide the signal length into different subintervals, and for each subinterval, the parameters are estimated by the IRNLS algorithm. With the segmentation, the system states in the last subinterval are passed into the next subinterval. The optimizations in different subintervals are conducted simultaneously and therefore reduce the errors of the initial state estimation.

Let the signal length $I = [0, T]$ be divided into $L \leq M$ subintervals $I_l (l = 1, 2, \ldots, L)$ such that each interval contains at least one measurement. The chaotic system has its own initial value $\mathbf{x}_0^l$ for each subinterval but shares the same parameter $\boldsymbol{\alpha}$. Denote the state of the chaotic system as $\mathbf{x}(t; \mathbf{x}_0^l, \bar{\boldsymbol{\alpha}})$ for $t \in I_l$. As a result, the problem (10) can be equally represented as

$$\begin{cases} \min_{\bar{\boldsymbol{\alpha}}, \mathbf{x}_0^2, \ldots, \mathbf{x}_0^L} \mathcal{L}_\lambda \left( \mathbf{x}_0^1, \mathbf{x}_0^2, \ldots, \mathbf{x}_0^L, \bar{\boldsymbol{\alpha}} \right) \\ \text{s.t.} \quad \mathbf{x}_0^l = \mathbf{x}\left( t_{l-1}^+; \mathbf{x}_0^{l-1}, \bar{\boldsymbol{\alpha}} \right) \quad (l = 2, \ldots, L) \\ \qquad \mathbf{x}_0^1 = \mathbf{x}_0 \end{cases} \quad (A.1)$$

where $t_l^+ = \max\{t : t \in I_l\}$ and

$$\mathcal{L}_\lambda \left( \mathbf{x}_0^1, \mathbf{x}_0^2, \ldots, \mathbf{x}_0^L, \bar{\boldsymbol{\alpha}} \right) = \sum_{l=1}^{L} \sum_{m: T_m \subset I_l} \left( y_m - \mathcal{H}_m \left( \mathbf{x}_0^l, \bar{\boldsymbol{\alpha}} \right) \right)^2 + \lambda \|\bar{\boldsymbol{\alpha}}\|_0 \quad (A.2)$$

The equality constraints in (A.1) enforce smoothness of the final trajectory and all discontinuities at the joins of the subintervals are eventually removed.

In the MS-IRNLS algorithm, $\|\bar{\boldsymbol{\alpha}}\|_0$ is approximated as the following form:

$$\|\bar{\boldsymbol{\alpha}}\|_0 \approx \tilde{\mathbf{W}} \left\| \mathbf{W}^{1/2} \bar{\boldsymbol{\alpha}} \right\|_2^2 \quad (A.3)$$

where $\tilde{\mathbf{W}} = diag(\tilde{\omega}_1, \tilde{\omega}_2, \ldots, \tilde{\omega}_N)$ is the outer weighted matrix, and $\mathbf{W} = diag(\omega_1, \omega_2, \ldots, \omega_N)$ is the inner weighted matrix. The former is updated by the outer iterative estimate of $\bar{\boldsymbol{\alpha}}$, denoted as $\tilde{\boldsymbol{\alpha}}$, with $\tilde{\omega}_i = \left( |\tilde{\alpha}_i| + \varepsilon \right)^{-1}$ ($1 \leq i \leq N$); and the latter is updated by the inner $j$ th iteration of $\bar{\boldsymbol{\alpha}}$, denote as $\tilde{\boldsymbol{\alpha}}^{(j)}$ with $\omega_i = \left( \left( \tilde{\alpha}_i^{(j)} \right)^2 + \varepsilon \right)^{-1/2}$. $\varepsilon$ is a small positive number to guarantee the non-negative definition of the diagonal matrix. We also define the maximum iteration number $J$ and the convergence error $err$ as the stopping criterion.

The algorithmic framework is described as follows:

**MS-IRNLS algorithm**

Step 1: Set the initial searching point $\tilde{\boldsymbol{\alpha}} = [\tilde{\alpha}_1, \tilde{\alpha}_2, \ldots, \tilde{\alpha}_N]^T$, $\tilde{\mathbf{x}}_0 = \left[ (\tilde{\mathbf{x}}_0^2)^T, (\tilde{\mathbf{x}}_0^3)^T, \ldots, (\tilde{\mathbf{x}}_0^N)^T \right]^T$, $\tilde{\mathbf{x}}_0^1 = \mathbf{x}_0$, and the parameters $J$, $\varepsilon$ and $err$.

Step 2: Compute the outer weighted matrix $\tilde{\mathbf{W}} = diag(\tilde{\omega}_1, \tilde{\omega}_2, \ldots, \tilde{\omega}_N)$, where $\tilde{\omega}_i = (|\tilde{\alpha}_i| + \varepsilon)^{-1}$. Let $\tilde{\boldsymbol{\alpha}}^{(0)} = \tilde{\boldsymbol{\alpha}}$, $\tilde{\mathbf{x}}_0^{(0)} = \tilde{\mathbf{x}}_0$, $j = 0$.

Step 3: Compute the inner weighted matrix $\mathbf{W}^{(j)} = diag(\omega_1, \omega_2, \ldots, \omega_N)$, where $\omega_i = \left( (\tilde{\alpha}_i^{(j)})^2 + \varepsilon \right)^{-1/2}$. Linearize the cost function and the equality constraints at the point $\boldsymbol{\theta}^{(j)} = \left[ (\tilde{\boldsymbol{\alpha}}^{(j)})^T, (\tilde{\mathbf{x}}_0^{(j)})^T \right]^T$ and solve the following problem:

$$\begin{cases} \min_{\boldsymbol{\theta}^{(j)}} \sum_{l=1}^{L} \sum_{m: T_m \in I_l} \left\| \mathbf{y}_m - \mathcal{H}_m\left( \tilde{\mathbf{x}}_0^{l(j)}, \tilde{\boldsymbol{\alpha}}^{(j)} \right) - \frac{\partial \mathcal{H}_m\left( \tilde{\mathbf{x}}_0^{l(j)}, \tilde{\boldsymbol{\alpha}}^{(j)} \right)}{\partial \boldsymbol{\theta}^{(j)}} \Delta \boldsymbol{\theta}^{(j)} \right\|_2^2 + \lambda \left\| (\tilde{\mathbf{W}} \mathbf{W}^{(j)})^{1/2} (\tilde{\boldsymbol{\alpha}}^{(j)} + \Delta \tilde{\boldsymbol{\alpha}}^{(j)}) \right\|_2^2 \\ s.t. \ \mathbf{x}\left(t_l^+; \tilde{\mathbf{x}}_0^{l(j)}, \tilde{\boldsymbol{\alpha}}^{(j)}\right) + \frac{\partial \mathbf{x}\left(t_l^+; \tilde{\mathbf{x}}_0^{l(j)}, \tilde{\boldsymbol{\alpha}}^{(j)}\right)}{\partial \boldsymbol{\theta}^{(j)}} \Delta \boldsymbol{\theta}^{(j)} = \tilde{\mathbf{x}}_0^{l+1(j)} + \Delta \tilde{\mathbf{x}}_0^{l+1(j)} \quad (l = 1, 2, \ldots, L-1) \end{cases}$$

Step 4: Update $\boldsymbol{\theta}^{(j+1)} = \boldsymbol{\theta}^{(j)} + \Delta \boldsymbol{\theta}^{(j)}$.

Step 5: When $j = J$ or $\left\| \tilde{\boldsymbol{\alpha}}^{(j+1)} - \tilde{\boldsymbol{\alpha}}^{(j)} \right\|_2 / \left\| \tilde{\boldsymbol{\alpha}}^{(j)} \right\|_2 \leq err$, go to Step 6; or else $j = j+1$, go to Step 3.

Step 6: When $\left\| \tilde{\boldsymbol{\alpha}}^{(j+1)} - \tilde{\boldsymbol{\alpha}} \right\|_2 / \left\| \tilde{\boldsymbol{\alpha}} \right\|_2 \leq err$, go to Step 7; or else $\tilde{\boldsymbol{\alpha}} = \tilde{\boldsymbol{\alpha}}^{(j+1)}$, go to Step 2.

Step 7: Output $\hat{\boldsymbol{\alpha}}_{out} = \tilde{\boldsymbol{\alpha}}^{(j+1)}$.

## Appendix B: Proof of the Theorem in Section IV.B

**Proof:** Firstly, we prove that the actual value $\boldsymbol{\alpha}$ is one of the local minima of the sparsity-regularized nonlinear least squares problem (10). Define the support set of $\boldsymbol{\alpha}$ as $\Omega$, i.e., $\Omega = \{k \mid \alpha_k \neq 0\} \subset \{1, 2, \cdots, B\}$. Then, we can derive from (8) that

$$\lim_{t \to 0} \mathcal{L}_\lambda (\mathbf{x}_0, \boldsymbol{\alpha} + t\mathbf{e}_k) = \begin{cases} \mathcal{L}_\lambda (\mathbf{x}_0, \boldsymbol{\alpha}) & k \in \Omega \\ \mathcal{L}_\lambda (\mathbf{x}_0, \boldsymbol{\alpha}) + 1 & k \notin \Omega \end{cases}$$

where $\mathbf{e}_k = [e_k(1), e_k(2), \cdots e_k(B)]^T$ with

$$e_k(i) = \begin{cases} 1 & i = k \\ 0 & i \neq k \end{cases}$$

Then there exists a neighborhood $B_\rho(\boldsymbol{\alpha})$ with radius $\rho$ such that $\mathcal{L}_\lambda(\mathbf{x}_0, \bar{\boldsymbol{\alpha}}) \geq \mathcal{L}_\lambda(\mathbf{x}_0, \boldsymbol{\alpha})$ for any $\bar{\boldsymbol{\alpha}} \in B_\rho(\boldsymbol{\alpha})$. Thus $\boldsymbol{\alpha}$ is a local minimum of problem (10).

Secondly, we prove that $\boldsymbol{\alpha}$ is the only local minimum in the neighborhood $B_\rho(\boldsymbol{\alpha})$ if $\mu_{\mathcal{S}^\lambda}(\mathbf{x}_0, \boldsymbol{\alpha}) < 1$. Assume that there exists another $\boldsymbol{\alpha}^* \in B_\rho(\boldsymbol{\alpha})$ satisfying $\mathcal{L}_\lambda(\mathbf{x}_0, \boldsymbol{\alpha}^*) = \mathcal{L}_\lambda(\mathbf{x}_0, \boldsymbol{\alpha})$, i.e., $\boldsymbol{\alpha}^*$ is also a local minimum in $B_\rho(\boldsymbol{\alpha})$. With the linearization of $\mathcal{L}_\lambda(\mathbf{x}_0, \boldsymbol{\alpha}^*)$ around $\boldsymbol{\alpha}$, $\mathcal{L}_\lambda(\mathbf{x}_0, \boldsymbol{\alpha}^*)$ can be approximately represented as

$$\mathcal{L}_\lambda(\mathbf{x}_0, \boldsymbol{\alpha}^*) = \mathcal{L}_\lambda(\mathbf{x}_0, \boldsymbol{\alpha}) + \mathcal{S}^\lambda(\mathbf{x}_0, \boldsymbol{\alpha}_0)(\boldsymbol{\alpha}^* - \boldsymbol{\alpha}) + o\left((\boldsymbol{\alpha}^* - \boldsymbol{\alpha})^2\right)$$

For small $\rho$ and omitting the high-order term, we can derive that

$$\mathcal{S}^\lambda(\mathbf{x}_0, \boldsymbol{\alpha})(\boldsymbol{\alpha}^* - \boldsymbol{\alpha}) = 0$$

Since $\boldsymbol{\alpha}^* \neq \boldsymbol{\alpha}$, the columns of the matrix $\mathcal{S}^\lambda(\mathbf{x}_0, \boldsymbol{\alpha})$ are linearly dependent. Thus $\mu_{\mathcal{S}^\lambda}(\mathbf{x}_0, \boldsymbol{\alpha}) = 1$, which is a contradiction to $\mu_{\mathcal{S}^\lambda}(\mathbf{x}_0, \boldsymbol{\alpha}) < 1$. Therefore, $\boldsymbol{\alpha}$ is the only local minimum in the neighborhood $B_\rho(\boldsymbol{\alpha})$.

As a result, there is only one solution $\boldsymbol{\alpha}$ to the sparsity-regularized nonlinear least squares problem (10) in the neighborhood $B_\rho(\boldsymbol{\alpha})$ if $\mu_{\mathcal{S}^\lambda}(\mathbf{x}_0, \boldsymbol{\alpha}) < 1$. Therefore, the sparsity-regularized nonlinear least squares problem is locally reconstructable. □

**Reference:**

Audoly, S., Bellu, G., D'Angio, L., Saccomani, M., P., & Cobelli, C. [2001] "Global